\documentclass[superscriptaddress,twocolumn,showpacs,preprintnumbers,amsmath,amssymb,pra]{revtex4}

\usepackage{graphicx}
\usepackage{dcolumn}
\usepackage{bm}

\newcommand{\ve}{\boldsymbol}
\newcommand{\tn}{\textnormal}

\newcommand{\D}{\tn{d}} 

\newcommand{\eq}{Eq.\,\eqref}

\begin{document}


\title{Ultrafast tilting of the dispersion of a photonic crystal \\ and adiabatic spectral compression of light pulses}

\author{Daryl M. Beggs}
\affiliation{FOM Institute AMOLF, Science Park 104, 1098 XG Amsterdam, The Netherlands} 
\affiliation{School of Physics and Astronomy, University of St. Andrews, St. Andrews, Fife, KY16 9SS, UK}
\author{Thomas F. Krauss}
\affiliation{School of Physics and Astronomy, University of St. Andrews, St. Andrews, Fife, KY16 9SS, UK}
\author{L. Kuipers}
\email{kuipers@amolf.nl} \affiliation{FOM Institute AMOLF, Science Park 104, 1098 XG Amsterdam, The Netherlands}
\author{Tobias Kampfrath}
\affiliation{FOM Institute AMOLF, Science Park 104, 1098 XG Amsterdam, The Netherlands} 
\affiliation{Fritz Haber Institute of the Max Planck Society, Faradayweg 4-6, 14195 Berlin, Germany} 

\date{\today}

\begin{abstract}

We demonstrate, by theory and experiment, the ultrafast tilting of the dispersion curve of a photonic-crystal waveguide following
the absorption of a femtosecond pump pulse. By shaping the pump-beam cross section with a nanometric shadow mask, different
waveguide eigenmodes acquire different spatial overlap with the perturbing pump, leading to a local flattening of the dispersion
by up to 11\,\%. We find that such partial mode perturbation can be used to adiabatically compress the spectrum of a light pulse
traveling through the waveguide.

\end{abstract}


\maketitle

Physical systems can often be described by their eigenmodes, that is, states of light or matter oscillating with a well-defined
frequency. For instance, the quantum-mechanical eigenstates of an electron spin are spin-up and spin-down~\cite{KittelBook},
while the optical eigenmodes of a photonic crystal are Bloch waves characterized by their wavevector $k$ and angular frequency
$\omega_k$~\cite{JoannopoulosBook}. Controlling the eigenfrequencies allows for the reversible manipulation of light and matter,
provided the dynamics evolve adiabatically, without exchange of energy between the eigenmodes. As an example, adiabatically
changing the frequency spacing of spin states is nowadays routinely employed to cool solids~\cite{KittelBook} or to control the
precession of a single spin~\cite{BerezovskyScience}. The state-dependent change in eigenfrequencies is realized by an external
magnetic or electric field which couples differently to spin-up and spin-down states.

Usually, however, {optical} eigenmodes (such as those of photonic crystals or microresonators) are perturbed by a homogeneous and
isotropic stimulus, leading to one and the same shift of all the
eigenfrequencies~\cite{MazurenkoPRL,VlasovNature,LinOE,TanabePRL,TobiasPRA,UphamAPEX}. A much higher level of control could be
achieved if one allows for a spatially varying perturbation that has different overlap with the light field of different modes.
Such a process would result in a modified frequency spacing and density of states~\cite{TanPRB}, which is particularly
interesting for a photonic crystal with its continuum of modes $k$. Figure 1(a) shows that a $k$-dependent mode-shift
$\Delta\omega_k$ would lead to a local tilting of the photonic dispersion $\omega_k$. A light pulse being a superposition of
these wavefunctions could then be spectrally compressed and temporally stretched, provided the dynamics were adiabatic. Moreover,
as first proposed by Yanik and Fan~\cite{YanikPRL}, such pulse could be also slowed down beyond the classical delay-bandwidth
limitation.

In this Letter, we demonstrate that perturbing selected regions of a photonic-crystal waveguide (PCW) by means of a spatially
shaped femtosecond laser pulse can locally flatten or steepen the dispersion curve. This novel procedure is akin to altering the
shape of the \lq\lq potential" confining the wavefunctions, thereby realizing a new dispersion relationship and density of
states~\cite{JoannopoulosBook}. Such partial mode perturbation (PMP) can be used to compress or expand the spectrum of a
picosecond light pulse traveling through the waveguide. As this operation proceeds adiabatically, it is reversible and features
high conversion efficiency.

\begin{figure}[b]\centering
\includegraphics[width=0.95\columnwidth]{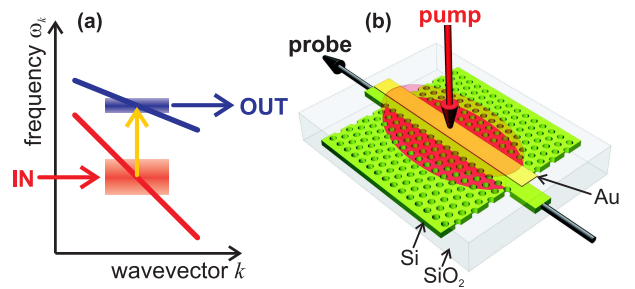}

\caption{(a)~Schematic of adiabatic light control. An input light pulse excites eigenmodes of a photonic crystal, whose
dispersion $\omega_k$ is then blueshifted and flattened by an external perturbation. When no light is scattered between
eigenmodes, the pulse undergoes an adiabatic spectral compression. (b)~Experimental realization. Whilst a probe pulse travels
through a Si-based photonic-crystal waveguide, a pump pulse generates free charge carriers in the illuminated Si regions. The
shadow mask causes a $k$-dependent spatial overlap between waveguide mode and pump pattern.}\label{fig:1}
\end{figure}

Our nanophotonic approach is schematically shown in Fig. 1(b). Whilst a light pulse travels through a PCW, a pump pulse incident
from above generates an electron-hole plasma in the Si parts of the structure, thereby reducing the refractive index. The index
change does not occur in the waveguide center as a shadow mask keeps pump light away from that region. Since PCW eigenmodes have
a strongly $k$-dependent lateral extent [Fig. 2(a)], each mode $k$ has a different spatial overlap with the pumped Si volume,
resulting in a pronounced variation of the mode-shift $\Delta\omega_k$ with respect to $k$. Note that the dynamics of the probe
pulse inside the waveguide will proceed adiabatically because the pump pulse respects all spatial symmetries of the
PCW~\cite{TobiasPRA,YanikPRL}. Breaking these symmetries would result in energy transfer between different
modes~\cite{DanielJOSAB} as has been shown in a complementary experiment~\cite{DongPRL}.

\emph{Theory.} We first explore our approach theoretically. To this end, we consider the frequency shift $\Delta\omega_k$ of mode
$k$ that results when the pump pulse alters the refractive-index landscape of the photonic crystal by an amount $f(\ve{r})\Delta
n$. Here, $0\leq f(\ve{r})\leq 1$ reflects the normalized spatial distribution of the absorbed pump energy, which induces a
change $\Delta n$ in the real part of the refractive index at positions $\ve{r}$ where $f(\ve{r})=1$. Perturbation
theory~\cite{JoannopoulosBook} then predicts a relative eigenfrequency shift
 \begin{equation}\label{eq:shift}
\frac{\Delta\omega_k}{\omega_k}=-\frac{\Delta n}{n}O_k
\end{equation}
which is directly proportional to the spatial overlap
 \begin{equation}\label{eq:Ok}
O_k=\int\D^3\ve{r}\,f(\ve{r})u_k(\ve{r})
\end{equation}
between the profiles of perturbation and eigenmode. In these relations, $n=3.48$ denotes the refractive index of the unpumped Si,
and $u_k(\ve{r})$ is the energy density of mode $k$ normalized according to $\int\D^3\ve{r}\,u_k(\ve{r}) = 1$.

In order to tilt the dispersion, $O_k$ needs to be $k$-dependent. More precisely, we have to realize a relative change $\Delta
v_k/v_k$ in the local slope $v_k=\partial\omega_k/\partial k$ of the dispersion curve that is much larger than the perturbation
$\Delta n/n$. Note that $\Delta v_k/v_k$ also quantifies the relative velocity change and relative spectral compression [Fig.
1(a)] of a light pulse encountering the dispersion change on-the-fly. In this respect, PCWs offer two unique benefits. First, the
lateral mode extent depends strongly on $k$ as seen in Fig. 2(a)~\cite{PetrovOE,LiamOE}. Second, using the slow-light modes with
low group velocity $v_k$ means that $\Delta v_k/v_k$ can be enhanced for a given $\Delta n/n$~\cite{TobiasPRA}.

\begin{figure*}\centering
\includegraphics[width=0.9\textwidth]{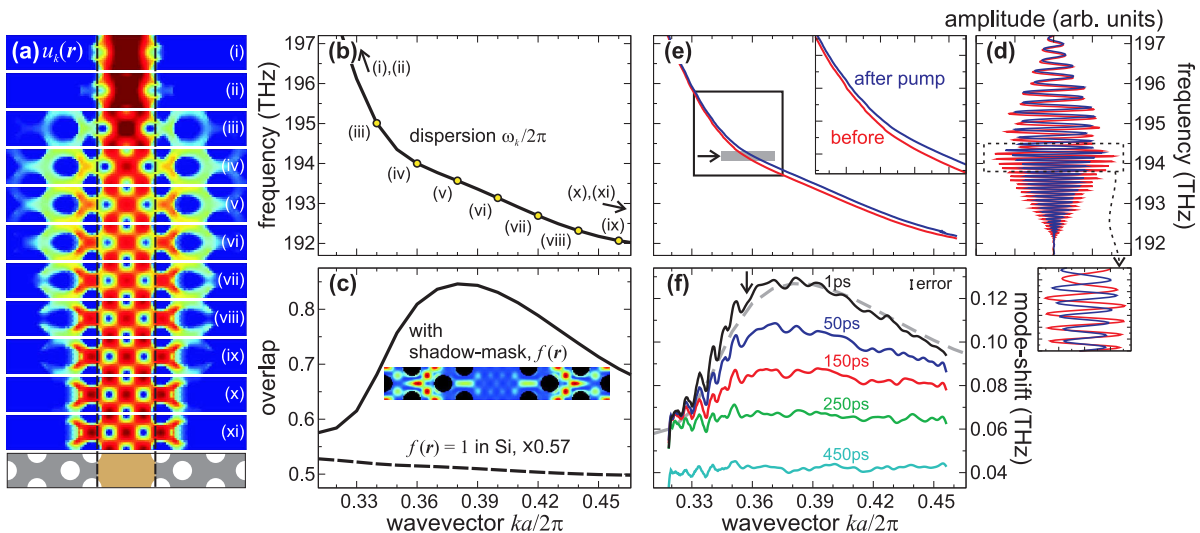}

\caption{Theoretical [(a)-(c)] and experimental [(d)-(f)] results for the waveguide dispersion changes. (a)~Bottom panel: shape
of unit cell of photonic-crystal waveguide together with shadow mask. Panels (i) to (xi): profiles of electromagnetic energy
density $u_k(\ve{r})$ of selected waveguide eigenmodes with $k$ indicated in (b). Note the logarithmic color scale.
(b)~Dispersion $\omega_k$ of the waveguide eigenfrequencies. (c)~Solid line: calculated overlap $O_k$ between the eigenmode $k$
[see (a)] and the pump pattern $f(\ve{r})$ given by the shadow mask (inset). Dashed line: $O_k$ for a spatially uniform change in
the Si refractive index.
(d)~Measured real part of the Fourier-transformed output pulse $E_\tn{o}(\omega,\tau)$ versus $\omega$ at 40\,ps before (red) and
1\,ps after (blue) waveguide pumping. (e)~Dispersion curves of unpumped and pumped waveguides as extracted from (d).
(f)~Dispersion shift $\Delta\omega_k$ at several delay times obtained by subtraction of the measured $\omega_k$ [see (e)]. Dashed
gray line: theoretical curve of Fig. 2(c), repeated for comparison. Bar: error estimated from signal variations at delays
$\tau<-15$\,ps.}\label{fig:2}
\end{figure*}

We calculated the eigenmodes~\cite{MPB} of a PCW [Fig. 1(b)], which can be pictured as a missing row of holes in a Si membrane
(thickness 220\,nm), perforated with a hexagonal pattern of holes (lattice period $a= 390$\,nm, hole diameter 220\,nm). The PCW
(length $L=180a=70.2\,\mu$m) has a 600-nm thick SiO$_2$ cladding on top of which a Au stripe (675\,nm wide, 20\,nm thick) serves
as a shadow mask. The overlap and, thus, the interaction between mask and the evanescent tail of light inside the waveguide are
negligible. Figure 2(a) displays the unit cell of the PCW and the profile $u_k(\ve{r})$ of the electromagnetic energy density for
several modes $k$. The respective values of $k$ can be inferred from the points labeled (i) to (xi) in the dispersion curve
$\omega_k$ of Fig. 2(b). With increasing $k$, the slope of $\omega_k$ and, thus, the group velocity decrease continuously, ending
up in a flat dispersion being characteristic of slow light ($v_k<c/50$ at $ka/2\pi>0.45$)~\cite{BabaNP}. As seen in Fig. 2(a),
the transition from fast to slow light is accompanied by drastic changes in the spatial mode profile~\cite{PetrovOE,LiamOE}.
Whereas the light field of modes (i) and (ii) lies nearly entirely under the shadow mask, it spreads out into the surrounding
lattice of holes for modes (iii) to (v) and then starts to contract again for modes (vi) to (xi).

The $k$-dependent lateral mode spread should be also reflected by the overlap $O_k$ of mode profile and pump pattern, as long as
the mask's shadow on the Si surface is not washed out by diffraction of the pump beam. Finite-difference time-domain
simulations~\cite{Rsoft} of the pump-pulse propagation indeed verify that $f(\ve{r})$ roughly follows the shape of the mask's
geometric shadow [inset of Fig. 2(c)]. The resulting overlap [\eq{eq:Ok}] of $f$ with the waveguide modes is shown in Fig. 2(c)
and exhibits the expected behavior: a steep rise for low wavevectors followed by a flatter decay at higher $k$. Thus, any pulse
populating these modes will be spectrally compressed or expanded upon the dynamic action of the perturbation. For comparison, the
dashed line shows the frequency shift for the case of a homogeneous perturbation profile ($f =1$ throughout the Si). The $k$
dependence of $O_k$ nearly vanishes because the total amount of mode energy inside the Si material is almost independent of $k$.

\emph{Experiment.} In order to put our PMP scheme to test, we fabricated a masked PCW with the same geometrical parameters used
in the calculations by means of electron-beam lithography and reactive ion etching. After filling with SiO$_2$, a Au shadow mask
is placed on top of the SiO$_2$ cladding by metal evaporation and lift-off. In order to excite the PCW, pump pulses (center
wavelength 810\,nm, duration 100\,fs full width at half maximum of the intensity, pulse energy 2\,nJ, repetition rate 80\,MHz)
from a Ti:sapphire laser pass a slit and cylindrical lenses resulting in a 3-$\mu$m thin line focus on the Si-membrane surface.
The slit ensures exclusive and homogeneous excitation along the length of the PCW. To measure the PCW transmittance over a large
bandwidth, a Fourier-limited probe pulse (1540\,nm, 180\,fs, 10\,pJ) from an optical parametric oscillator is coupled into the
waveguide at a delay $\tau$ after excitation by the pump pulse. We pick up the probe at the output facet and determine the
complex-valued Fourier amplitude $E_\tn{o}(\omega,\tau)$ of its electric field by means of spectral
interferometry~\cite{LepetitJOSAB,TobiasOL}. By varying the delay between pump and probe pulse, we obtain a two-dimensional data
set $E_\tn{o}(\omega,\tau)$.

Figure 2(d) shows the real part of the measured $E_\tn{o}(\omega,\tau)$ at 40\,ps before PCW pumping ($\tau=-40$\,ps, red line)
and at 1\,ps after ($\tau=1$\,ps, blue line). In both cases, two regimes with slow and fast oscillations are observed. Assuming
single-mode propagation, the phase acquired by the probe pulse after propagation through the PCW equals $k(\omega)L$. Thus,
$\tn{Re}\,E_\tn{o}=|E_\tn{o}|\cos(kL)$ versus $\omega$ oscillates more rapidly for slower light as the inverse group velocity
$\partial {k}/\partial\omega$ is larger. Therefore, the fast and slow oscillations in Fig. 2(d) are signatures of slow and fast
light, respectively. Solving for the phase of $E_\tn{o}$ allows us to extract the wavevector $k$ as a function of $\omega$ (apart
from an unknown offset since only phase differences can be measured). The so-obtained waveguide dispersion $\omega_k$ is shown in
Fig. 2(e) before and after Si pumping. One clearly recognizes the regions of fast and slow light, and the shape of both curves is
in very good agreement with that of the one calculated [Fig. 2(b)].

A superficial glance at Fig. 2(e) might lead to the notion that the dispersion directly after PCW pumping ($\tau= 1$\,ps) is just
a blueshifted version of the dispersion of the unpumped waveguide. However, a magnified view [inset of Fig. 2(e)] reveals
departures from such rigid-shift-type behavior. This observation becomes even more apparent when we subtract one of the
dispersion curves from the other. Figure 2(f) presents the resulting pump-induced mode-shift $\Delta\omega_k$ at various delays
after PCW pumping. Shortly after excitation ($\tau= 1$\,ps), the mode-shift varies strongly with the wavevector, from
$\Delta\omega_k/2\pi=0.060$\,THz (at $ka/2\pi = 0.33$) to over 0.120\,THz (0.38), before dropping back to 0.095\,THz (0.45). This
curve agrees excellently with the theoretical prediction [solid line in Fig. 2(c), gray dashed line in Fig. 2(f)]. Comparison of
the calculated overlap $O_k$ and the measured mode-shift $\Delta\omega_k$ with \eq{eq:shift} allows us to estimate the
pump-induced change in the Si refractive index to $\Delta n/n\approx -1.5\cdot 10^{-3}$. Interestingly, at $ka/2\pi = 0.37$
[arrows in Figs. 2(e) and 2(f)], the measured $\Delta\omega_k$ and, thus, $O_k$ have a quite steep slope whereas the slope of
$\omega_k$ is rather flat, resulting in a measured ratio $|\Delta v_k/ v_k|$ of as much as $12\,\%$. This value is more than 80
times larger than the perturbation $|\Delta n/n|$, showing that our nanophotonic approach [Fig. 1(b)] indeed causes a strongly
$k$-dependent frequency shift of the photonic-crystal eigenmodes.

At later times after pump excitation, charge-carrier diffusion and recombination in Si are expected to modify the
refractive-index distribution $f(\ve{r})\Delta n$~\cite{TanPRB}. Indeed, with increasing delay $\tau$, the measured
$\Delta\omega_k$ [Fig. 2(f)] undergoes an overall decrease and returns into a flat line ($\tau\geq 250$\,ps). This final,
virtually $k$-independent mode-shift agrees well with the calculated $\Delta\omega_k$ for a homogeneous excitation pattern
[dashed line in Fig. 2(c)]. Thus, after about 250\,ps, the pump-induced charge carriers have diffused into the Si underneath the
675\,nm wide shadow mask, resulting in a homogenous carrier density. These numbers allow us to estimate a mean diffusion constant
of $\sim 10\,\tn{cm}^2\,\tn{s}^{-1}$, which is in good agreement with previous measurements~\cite{LiPRB}. Whereas the observed
flattening of the dispersion change arises from carrier diffusion, the temporal decrease of the $k$-averaged $\Delta\omega_k$ on
a time scale of about 500\,ps is a consequence of carrier recombination, predominantly taking place at the surface of the
PCW~\cite{TanPRB}.

\begin{figure}[t]\centering
\includegraphics[width=1\columnwidth]{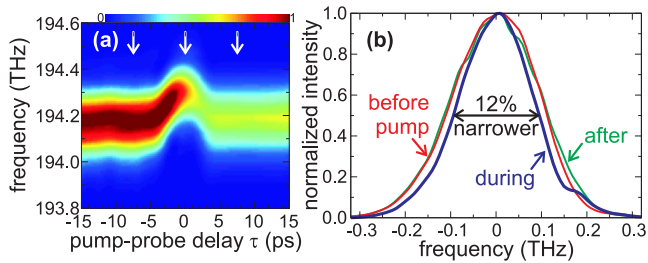}

\caption{Adiabatic spectral compression of a light pulse. (a)~Output intensity spectra versus $\tau$ for a Fourier-limited input
pulse of frequency 194.2\,THz and duration of 1.9\,ps. (b)~Output intensity spectra at delays indicated by arrows in (a), that
is, for the mode-shift occurring long before, after, or while the probe is inside the waveguide. Spectra are normalized to same
height and shifted to have center frequency $\omega=0$.}\label{fig:3}
\end{figure}

\emph{Spectral compression.} We finally illustrate how the ultrafast modifications of the PCW dispersion can be used to
spectrally compress a light pulse. For this purpose, we consider a Fourier-limited Gaussian pulse with a duration of 1.9\,ps and
a center frequency of 194.2\,THz as PCW input [arrows in Figs. 2(e) and 2(f)] for the following two reasons. First, the spectrum
coincides with a part of the dispersion that undergoes a significant pump-induced flattening $\Delta v_k/v_k$ (see above).
Second, this pulse allows one to confine 96\,\% of its energy within the length of the waveguide as its duration is significantly
shorter than its transit time of 5.6\,ps.

We are able to determine the PCW response to the fictitious 1.9-ps input pulse from the two-dimensional data set
$E_\tn{o}(\omega,\tau)$ [Fig. 2(d)] that was measured using the much more broadband 180-fs probe pulse. This
approach~\cite{TobiasOL,TobiasPRA} merely presumes that the waveguide response to the weak probe pulse is linear, which we have
verified experimentally.
The extracted output spectrum is shown in Fig. 3(a) as a function of the delay between pump and probe pulse. At delays
$\tau<-5$\,ps and $\tau>5$\,ps, no dynamics occur as the probe pulse encounters a fully unexcited or excited waveguide,
respectively. The energy of the output pulse is smaller in the excited case because free charge carriers lead to additional light
absorption. At $\tau=0$, the probe pulse is fully contained in the PCW when the pump-induced shift $\Delta\omega_k$ of the
dispersion curve occurs. As the light populating these modes is frequency-shifted as well [Fig. 1(a)], we find a clear blueshift
of the PCW output around $\tau=0$ [Fig. 3(a)]. The center frequency increases by 0.12\,THz, in good agreement with the magnitude
of the measured $\Delta\omega_k$ [Fig. 2(f)].

In order to evaluate further spectral modifications of the blueshifted pulse, Fig. 3(b) displays the extracted output spectra
before and after waveguide excitation as well as at $\tau=0$ [see arrows in Fig. 3(a)]. For better comparison, the spectra are
normalized to the same height, and their center frequencies are shifted to frequency $\omega=0$. Note that the spectrum of the
blueshifted pulse is noticeably narrower than the spectra obtained before and after PCW pumping. The relative spectral decrease
of the full width at half maximum amounts to about 11\,\%, which agrees well with the relative change $\Delta v_k/v_k$ of 12\,\%
of the slope of the dispersion curve [Figs. 2(e) and 2(f)]. Therefore, our findings are consistent with the adiabatic spectral
compression of light as anticipated in Fig. 1(a). The conversion efficiency is better than 60\,\% and only limited by
free-carrier absorption. We note that adiabatic spectral expansion can also be obtained when pulses in the slow-light region at
193.3\,THz [Fig. 2(e)] are used (data not shown).

In conclusion, we have demonstrated that partial mode perturbation can be used for all-optical adiabatic pulse-bandwidth
compression.
This process is significant because it overcomes the fundamental bandwidth-delay constraint in optics~\cite{YanikPRL}.
Ultimately, such a process can generate arbitrarily small group velocities for any light pulse with a given bandwidth, without
the need for intrinsic material resonances. The unavoidable signal loss due to free-carrier absorption could be prevented by
using tuning mechanisms like the instantaneous optical Kerr effect. Besides the dynamical slow-down of light, spectral
compression could also find application as a magnifying time lens~\cite{KolnerOL,BiancalanaPRE,FosterNature}: the uniform
shrinking of the frequency axis by a factor $1+\Delta v_k/v_k$ yields an expansion of the pulse by the inverse factor in the time
domain. The magnitude of the spectral compression could be enhanced further by optimizing the profiles of the perturbing pump or
by modifying the photonic crystal itself.
One could repeat the spectral compression in cascaded photonic crystals: the smaller the bandwidth of the pulse becomes, the
smaller the slope of the dispersion of the next waveguide-stage can be chosen, leading to a larger relative compression $\Delta
v_k/v_k$.


We acknowledge funding through the EU FP6-FET \lq\lq SPLASH" project. This work is also part of the research program of FOM,
which is financially supported by the NWO.


\begin{thebibliography}{99}
\bibitem{Rsoft}

We used the commercially available software Fullwave by Rsoft.
\end{thebibliography}
\end{document}